\documentclass[12pt]{iopart}
\usepackage{graphicx}
\begin{document}

\title[Phase transitions with finite atom number]{Phase transitions with finite atom number in the Dicke Model}

\author{J. G. Hirsch, O. Casta\~nos, E. Nahmad-Achar, and R. L\'opez-Pe\~na}

\address{Instituto de Ciencias Nucleares,
Universidad Nacional Aut\'onoma de M\'exico 
Apdo. Postal 70-543, Mexico D. F., C.P. 04510}
\ead{hirsch@nucleares.unam.mx}
\begin{abstract}

Two-level atoms interacting with a one mode cavity field at zero temperature have order parameters which reflect the
presence of a quantum phase transition at a critical value of the atom-cavity coupling strength. 
Two popular examples are the number of photons inside the cavity and the number of excited atoms.
Coherent states provide a mean field description, which becomes exact in the thermodynamic limit.
Employing symmetry adapted (SA) SU(2) coherent states (SACS) the critical behavior can be described for a finite number of atoms.
A variation after projection treatment, involving a numerical minimization of the SA energy surface, associates the finite number phase transition with a discontinuity in the order parameters, which originates from a competition between two local minima in the SA energy surface.
\end{abstract}
%

\section{Introduction}

Cavity quantum electrodynamics describes a system containing elements which can absorb and emit radiation interacting with quantized radiation. The simplest example is that of one atom interacting with a single cavity mode, which exhibits the presence of quantum revivals and squeezing \cite{eberly,kulinski}. For $N$ non-interacting two-level atoms \cite{dicke}, above some critical coupling strength, all atoms are predicted to absorb and emit light collectively and coherently, in a superradiant mode \cite{hepp}. This mode has been observed in a driven system \cite{baumann} which, under certain assumptions, is described by the same dynamics \cite{nagy}. An analytical description, based in a truncated Holstein-Primakoff expansion and valid in the thermodynamic limit, when the number of atoms goes to infinity, has been widely used \cite{nagy,Ema03,Dim07,Nat10}, although it is divergent at the critical point. Finite size corrections have been studied in Ref. \cite{Vid06}. A similar behavior is found in superconducting circuit QED systems  \cite{Nat10}, where the number of elements is small and it is important to take explicitly into account these finite size effects.

The description of the phase transition at zero temperature and for a large number of atoms can be performed analytically through the mean field formalism. The radiation is described with harmonic oscillator, Heisenberg-Weyl (HW) coherent states (CS) \cite{meystre,manko}. The atomic sector can be described with  HW-CS, employing the Holstein-Primakoff realization of the angular momentum operators \cite{Ema03,Hir12}, or with spin SU(2) CS \cite{gilmore1972}. In both cases the energy per atom, the number of photons per atom and the fraction of excited atoms are very well described in the large $N$ limit, while other observables are misrepresented due to the violation of the Hamiltonian parity symmetry in the description of the ground state wave function in the superradiant phase. This difficulty can be overcome by employing symmetry adapted coherent states (SACS), which by construction restore the parity symmetry \cite{Cas05,Cas06,ocasta2}.  

While SACS allows for a better description of the ground and first excited states than CS, the energy surface obtained with SACS is more complicated, preventing a direct analytical treatment of the critical points. Employing the CS expressions obtained for them in the SACS description provides wave functions which have good overlap with the exact wave function calculated at finite $N$ (but for a small vicinity around the phase transition), 
allowing to express the expectation values of matter and field observables with elegant analytic forms in the superradiant regime \cite{Cas11a,Cas11b,Cas10,Rom12,Cal12}. Many of them coincide with the mean field ones for large $N$, while some others, noticeably the fluctuations, differ. This analytic description fails, however, to describe effects of order $1/N$ in the normal phase at finite $N$, which are absent in the mean field description.

To go beyond this point, we have performed a variational treatment of the SACS energy surface. In what follows we present the details of this formalism, which exhibits a phase transition at finite N at values of the coupling constant which are larger than those found in the thermodynamic limit, and which has non-trivial values of the order parameters, in agreement with the numerical calculations. These results can be applied to the description of the phase transition at large but finite N, where numerical simulations are difficult. 

\section{The Dicke Model}

The Dicke Hamiltonian allows the description of $N$ two-level atoms, with energy separation $\omega_A$, interacting collectively with a one mode radiation field, whose frequency (times $\hbar$) is employed as the unit of energy along this work. Photon creation and annihilation operators are represented by the operators $a^{\dagger}$ and $a$, the atomic relative population operator by $J_{z}$, and the atomic transition operators by $J_{\pm}$.  The Hamiltonian reads 

	\begin{equation}
		H_{D}= a^{\dagger}a + \omega_{A}\,J_{z}
		+\frac{\gamma}{\sqrt{N}}\left(a^{\dagger} + a \right) \left( \,J_{-}
		+ \,J_{+}\right)\ ,
		\label{D}
	\end{equation}
where  $\gamma $ is the (adimensional) atom-field coupling parameter.

The eigenvalues $j$ associated with the amplitude of angular momentum operator ${\hat J}^2$ are fixed as $j=N/2$ along this work, where we are restricting our analysis to the fully symmetric representation of the atomic states, which are enough to describe its collective behavior. When there is no interaction the number of excited atoms is $j+m$, where $m$ is the eigenvalue of ${\hat J}_z$; the number of photons inside the cavity is $\nu$, the eigenvalue of the photon number operator $a^{\dagger} a$; and the excitation number is $\lambda = j +m + \nu$, the eigenvalue of the excitation number operator $\hat\Lambda$.
$\lambda$ is a conserved quantity in the Jaynes- and Tavis-Cummings models \cite{jaynes,tavis}, where the rotating wave approximation is employed \cite{ocasta2,ocasta1}, but the counterrotating terms in the Dicke Hamiltonian ($a^{\dagger}  \,J_{+}, a \,J_{-}$) mix them. This has far reaching consequences: there are no finite subspaces where the Dicke Hamiltonian can be exactly diagonalized, it is not integrable, and around the phase transition it exhibits the presence of quantum chaos  \cite{Ema03a,Lam04}.

There is, however, a remnant symmetry: the parity symmetry  operator $e^{i \pi \hat{\Lambda}}$ commutes with the Dicke Hamiltonian, allowing the classification of its eigenstates, as they belong to the  symmetric or to the antisymmetric representations of the cyclic group $C_2$ \cite{Cas11a,Cas11b,Cas10}.

\section{The mean field analysis}

As mentioned in the introduction, the mean field treatment allows for the appropriate description of many intensive observables in the thermodynamic limit. In this section we briefly review the main results, which have already been reported \cite{Cas11a,Cas11b}. The starting point is a variational state  built as a direct product  Heisenberg-Weyl coherent states $\vert\alpha\rangle$~\cite{meystre,manko}  for the photon sector and $SU(2)$ or spin coherent states $\vert\zeta\rangle$~\cite{gilmore1972,Cas05} for the particle sector, i.e., $\vert\alpha, \,\zeta\rangle=\vert\alpha\rangle \otimes\vert\zeta\rangle$. They are 
\begin{eqnarray}
		\vert\alpha \rangle &= \left( \exp\left|\alpha\right| \right)^{-1} \exp(\alpha a^{\dagger}) \vert 0\rangle =
		\left( \exp\left|\alpha\right| \right)^{-1}
		\sum\limits_{\nu=0}^{\infty}
		\frac{\alpha^{\nu}}{\sqrt{\nu!}}\,| \nu\rangle ,\nonumber \\
		\vert \zeta\rangle &=  \left(1+\left|\zeta\right|^{2}\right)^{-N/2} \exp(\zeta J_+) \vert j, \, -j\rangle
		\\ & =
		\left(1+\left|\zeta\right|^{2}\right)^{-N/2}
		\sum\limits_{m=-j}^{j}\,
		\left( \begin{array}{c} 2j \\ j + m \end{array} \right)^{1/2} 
		\zeta^{j + m}\,\vert j,\,m\rangle \ , \nonumber
	\label{trial}
	\end{eqnarray}
where the ket $\vert \nu \rangle$ is an eigenstate of the photon number operator, and  $\vert j,\, m \rangle$ is a Dicke state. This trial state describes $N=2j$ particles distributed in all the possible ways between the two levels, and up to an infinite number of photons in the cavity. 

Expressing the internal variable $\alpha$ in terms of the quadratures $(q,p)$ associated with harmonic oscillator realization of the field, and $\zeta$ in terms of  the angles  $(\theta,\phi)$ which determine its stereographic projection on the Bloch sphere,
	\begin{equation}
		\alpha=\frac{1}{\sqrt{2}}\left(q+i\,p\right)\ ,\qquad
		\zeta = \tan\left(\frac{\theta}{2}\right)\,\exp\left(-i\,\phi\right)\ ,
		\label{dzeta}
	\end{equation}
the expectation value of the Hamiltonian with respect to the trial states is obtained. From it, the energy surface is defined dividing by the number of particles. It is
	\begin{eqnarray}
		E(q,\,p,\,\theta,\,\phi)= 
\frac{1}{2 N} \left( p^2 + q^2 \right) - \frac{1}{2 } \omega_A \cos\theta 
+ \frac{\sqrt 2 \gamma }{\sqrt N} q \, \sin\theta \cos\phi \ ,
\label{emf}
	\end{eqnarray}
It depends on two control parameters ($\omega,\gamma$) and four order parameters ($q, p, \theta, \phi$). 

The catastrophe formalism~\cite{gilmore3}  allows the obtention of the critical points $q_c,\,p_c,\,\theta_c,\,\phi_c$ and the associated separatrix, the geometric place in the control parameter space where a catastrophe, {\em i.e.} a sudden change in the ground state of the system, reflected in the order parameters, takes place. They are the solutions of the system of equations
\begin{equation}
\frac{\partial E}{\partial u} = 0, ~~~~~~~u = q,\,p,\,\theta,\,\phi.
\label{der}
\end{equation}
Those which correspond to minima in the energy surface are \cite{Cas11a,Cas11b}
\begin{equation}\label{criticos}
		\begin{array}{lllll}
		\theta_{c}=0\,,& q_{c}=0\, ,&p_{c}=0\, ,&\hbox{
                for } \vert\gamma\vert < \gamma_c \, ,\\
		\theta_{c}=\arccos(\gamma_c/\gamma)^{2}
		\, ,&q_{c}=-2\,\sqrt{j}\,\gamma\,
		\sqrt{1-(\gamma_c/\gamma)^{4}} 
		\cos{\phi_c}\, ,&p_{c}=0\, ,&\hbox{ for
                }\vert\gamma\vert > \gamma_c\ .
		\end{array} 
	\end{equation}
The condition $ \vert\gamma\vert < \gamma_c$ defines the normal phase, where the ground state has zero photons and no excited atoms. The superradiant phase, described in the second row, assumes  $\omega_{A}  > 0$.
Introducing the variable $x= \gamma/\gamma_c$, the energy at the minima takes the values \cite{Cas11b}
	\begin{equation}
	    E_{\hbox{normal}}=-2N\,\gamma_{c}^{2}\ ,\quad
                E_{\hbox{superradiant}}=-N\,\gamma_{c}^{2}\,x^{2}\,\left(
                1+x^{-4}\right)\ .
                \label{ecoherente}
	\end{equation}

\section{Symmetry-Adapted Coherent States (SACS)}

The variational states which preserve the symmetry of the Dicke
Hamiltonian can be built from even and odd parity coherent
states~$\vert\alpha\rangle\otimes\vert\zeta\rangle$.  They have the form \cite{Cas11a,Cas11b}
	\begin{equation}
		|\alpha,\,\zeta \rangle_{\pm}={\cal N}_{\pm}\Big(
		\vert\alpha\rangle\otimes\vert\zeta\rangle\pm\
		\vert-\alpha\rangle\otimes\vert-\zeta\rangle\Big)\ ,
		\label{sacs}
	\end{equation}
with normalization ${\cal N}_{\pm}$ 
       \begin{equation}
          {\cal N}_{\pm}^{-2}=2\,\left(1\pm\exp\left(-2\,|\alpha|^{2}
          \right)\left(\frac{1-|\zeta|^2}{1+|\zeta|^2}\right)^{N}\right)\ .
          \label{nsacs}
       \end{equation}
The above expressions can be used to evaluate the expectation value of the Dicke Hamiltonian. When divided by the number of particles, it represents the energy surface with respect to the symmetry adapted states (SAES), which has the form \cite{Cas11b}
	\begin{eqnarray}
	E_\pm &=& \pm \frac{1}{2} \left(p^2+q^2\right)
	\left\{1-\frac{2}{1 \pm e^{\pm(p^2+q^2)} (\cos\theta)^{\mp
            N}}\right\}\nonumber \\
	&-&\frac{N}{2} \, \omega_{A}
	 \left\{(\cos \theta)^{\pm 1} 
	 \pm \frac{\tan^2\theta \, \cos\theta }{1 \pm e^{\pm(p^2+q^2)} 
	(\cos\theta)^{\mp N} }\right\} \nonumber \\
	&+& \sqrt{2 \, N} \, \gamma  \left\{\frac{\pm p \, \tan\theta \, 
	\sin\phi + q \, e^{p^2+q^2}
	\sin\theta \, \cos \phi \, (\cos\theta)^{-N} }{
	e^{p^2+q^2} (\cos\theta)^{-N}  \pm 1 } \right\} \, .
	\label{symad}
	\end{eqnarray}
In the limit $N \rightarrow \infty$, $E_+$ reduces to \Eref{emf},
when $\vert\cos\theta\vert\neq 1$. However, it is important to stress
that the analysis can be carried out {\em for any value of} $N$, and in this
contribution we work at {\em finite} $N$.

In our previous works we have used the SACS, \Eref{sacs}, with the critical points of the standard CS given in \eref{criticos}. It allowed us to obtain analytical expressions for all the observables of interest \cite{Cas11b}. In the limit $N \rightarrow \infty$ they coincide with those obtained with the coherent state,  with the exception of $(\Delta\hat{q})^{2}$ and $(\Delta\hat{J}_{x})^{2}$. These two exceptions arise because $\langle\hat{q}\rangle=\langle\hat{J}_{x}\rangle=0$ for the SACS but not for the CS. As expected, the SACS results are in close agreement with the numerical calculations for finite $N$ \cite{Cas11a}. 

There are, however, clear limitations in this mixed approach, where the coherent states have the parity symmetry restored but the critical points are not obtained by minimizing the corresponding energy surface. In the normal phase the solutions are trivial but not correct for finite $N$, because effects of order $1/N$ are not considered at the mean field level. This problem extends into the superradiant phase: in a vicinity of the phase transition the functions employed do not provide a minima of the energy surface given in \Eref{symad}, because they are not solutions of \eref{der} \cite{Cas11b}. 

\section{Full variational treatment}

The exact critical points of the SAES \eref{symad} are obtained by applying to it the conditions given in \eref{der}, {\it i.e.} its first derivatives with respect to the four internal variables should cancel. The partial derivative with respect to $p$ implies $p=0$. Using it, the partial derivative with respect to $\phi$ leads to $\sin \phi = 0$. For $\omega_A > 0$, the minima occurs for  $\phi = 0$. Introducing $z = \cos \theta $ the other two equations which must be satisfied, for the  even parity case, are:

\begin{eqnarray}
\frac{\partial E_+}{\partial q} = &\frac{1}{z}e^{-2 q^2} \left[-q z^{1+4 j}+e^{2 q^2} z \left(q+2 \sqrt{j-j z^2} \gamma \right) \right. 
\label{ddq}\\
&  \left. +2 e^{q^2} z^{2 j} \left(q^3 z+z \sqrt{j-j z^2} \gamma +2 q^2 z \sqrt{j-j z^2} \gamma -j q \left(-1+z^2\right) \omega_A \right)\right] = 0,
\nonumber\\
\frac{\partial E_+}{\partial \theta} = &
\frac{1}{z^2}e^{-2 q^2} \left[-j z^{4 j} \sqrt{1-z^2} \omega_A +e^{2 q^2} \left(2 \sqrt{j} q z^3 \gamma +j z^2 \sqrt{1-z^2} \omega_A \right) \right.
\label{ddt} \\
&  +e^{q^2} \sqrt{j} z^{2 j} \left(2 q z^3 \gamma -4 j q z \left(-1+z^2\right) \gamma +2 j^{3/2} \left(1-z^2\right)^{3/2} \omega_A \right. \nonumber\\
&\left. \left. +\sqrt{j} \sqrt{1-z^2} \left(2 q^2 z+\left(-1+z^2\right) \omega_A \right)\right)\right] = 0. \nonumber
\end{eqnarray}

We have explored two ways to find the minima numerically: solving the above set of coupled non-linear differential equations, and exploring the SAES to find them. Employing {\em Mathematica} \cite{Math}, we found that the most efficient way was the second one. In the following figures the symmetry adapted energy surface is plotted for five values of the interaction strength $\gamma$, at resonance ($\omega_A = 1$). On the left hand plots, the horizontal axis represents $q$, and the vertical axis $\theta$. Continuous lines connect regions with equal energies, and the minima are shown with large dots

\begin{figure}
\begin{tabular}{ccc} 
\begin{tabular}{c}  $\gamma = 0.545$ \\  ~\\ ~ \\ ~ \\  ~\\ ~ \\ ~    \end{tabular}
 &     \includegraphics[width = 0.4\textwidth]{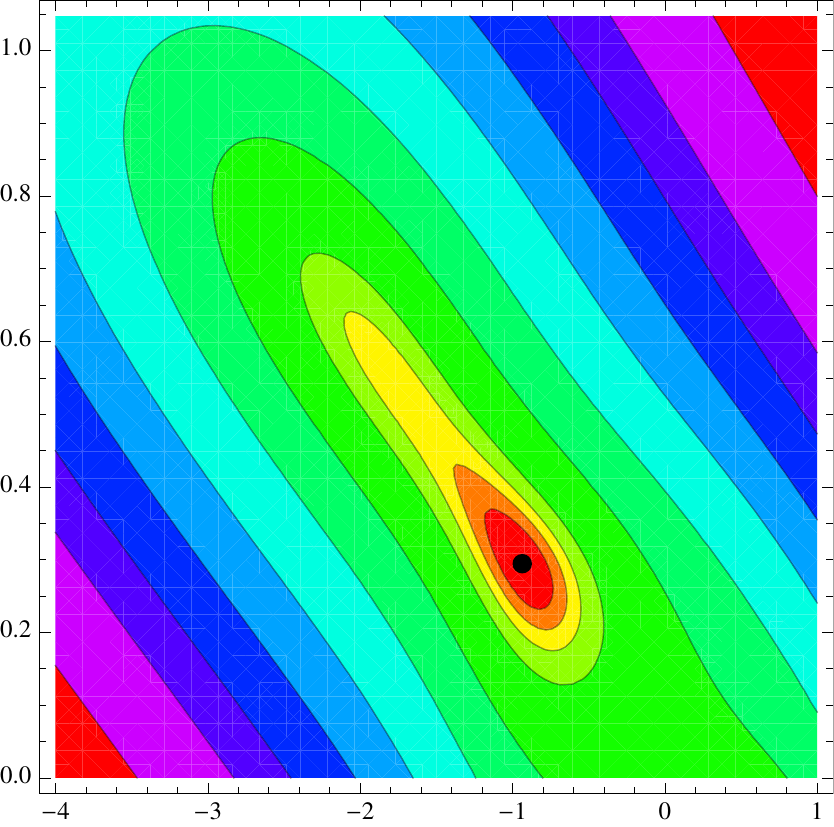} &  
        \includegraphics[width = 0.4\textwidth]{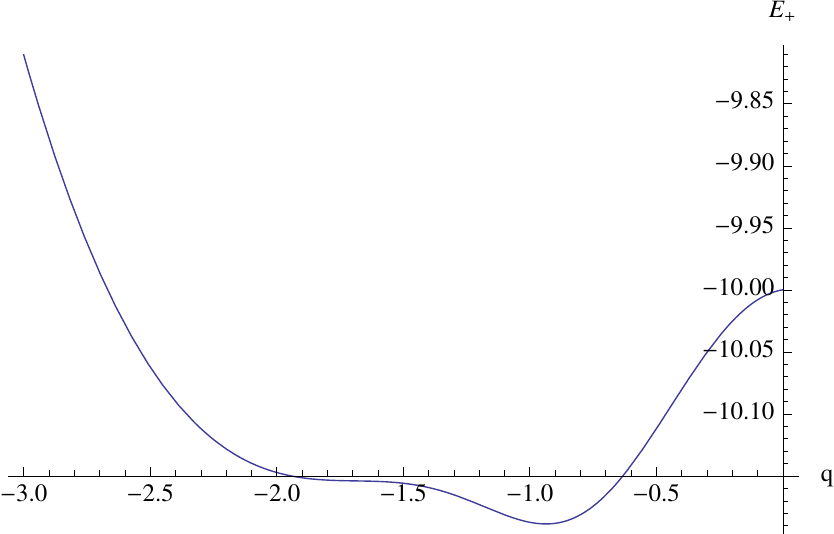}  \\  
\begin{tabular}{c}  $\gamma = 0.550$ \\  ~\\ ~ \\ ~ \\  ~\\ ~ \\ ~    \end{tabular} &           \includegraphics[width = 0.4\textwidth]{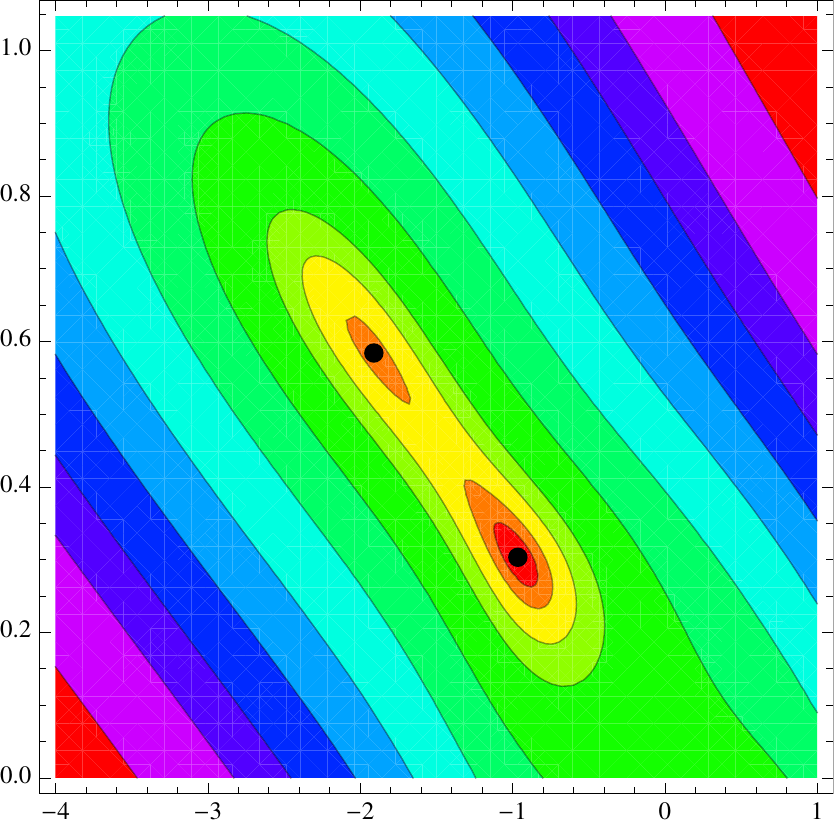} &  
        \includegraphics[width = 0.4\textwidth]{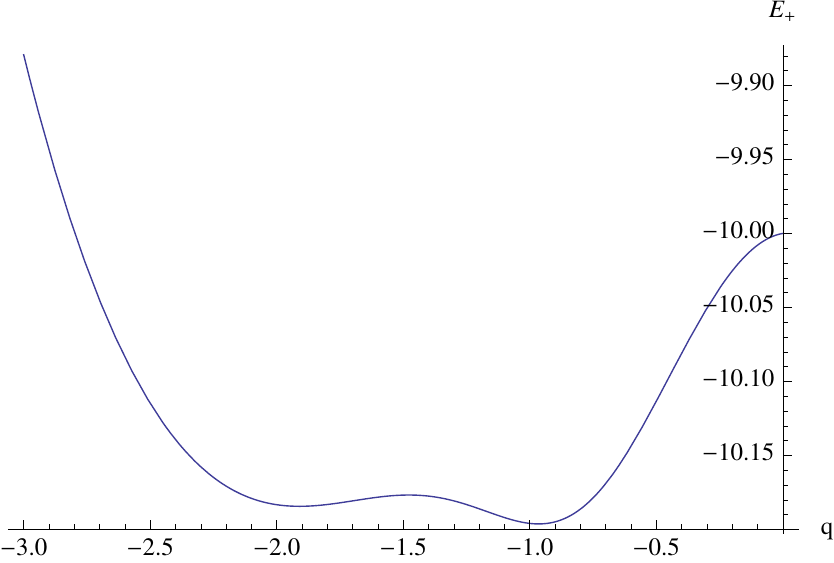}    \\ 
\begin{tabular}{c}  $\gamma = 0.552$ \\  ~\\ ~ \\ ~ \\  ~\\ ~ \\ ~    \end{tabular}&           \includegraphics[width = 0.4\textwidth]{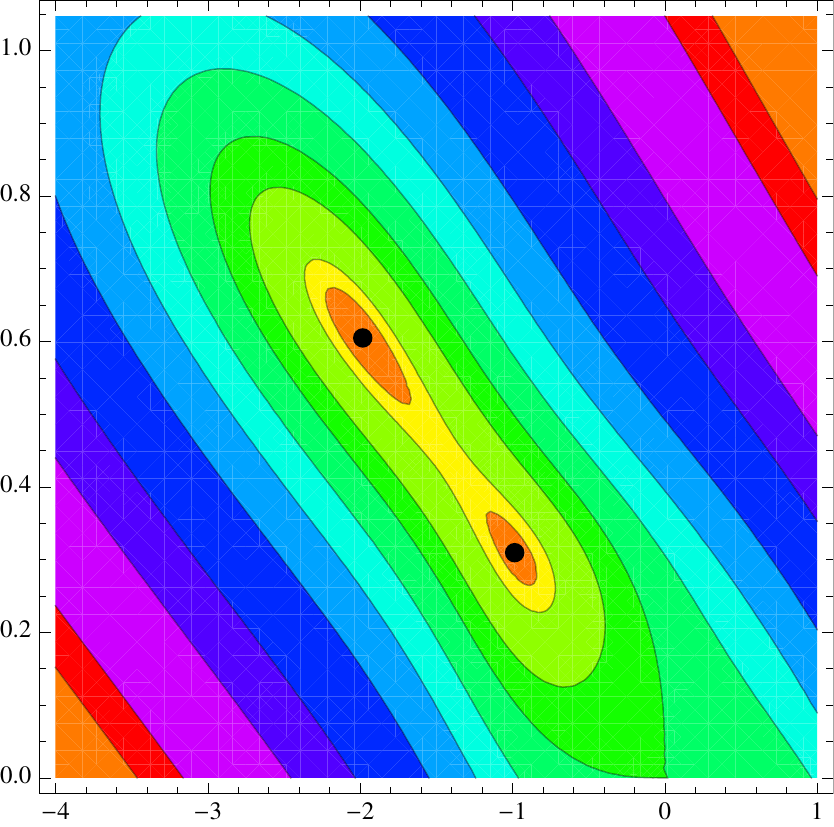} &  
        \includegraphics[width = 0.4\textwidth]{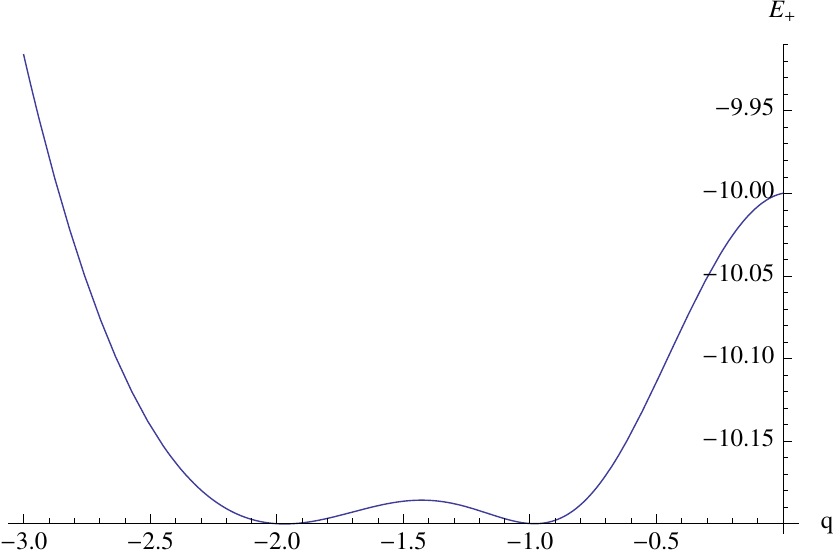}  
\end{tabular}
\vskip -1.5cm
\caption{The symmetry adapted energy surface showing the minima for $\gamma= 0.545, 0.550, 0.552$ plotted as a function of $q$ and $\theta$ (left) and as function of $q$ (right).}
\label{fig1}
\end{figure}
\begin{figure}
\begin{tabular}{ccc} 
 \begin{tabular}{c}  $\gamma = 0.555$ \\  ~\\ ~ \\ ~ \\  ~\\ ~ \\ ~    \end{tabular}      &     \includegraphics[width = 0.4\textwidth]{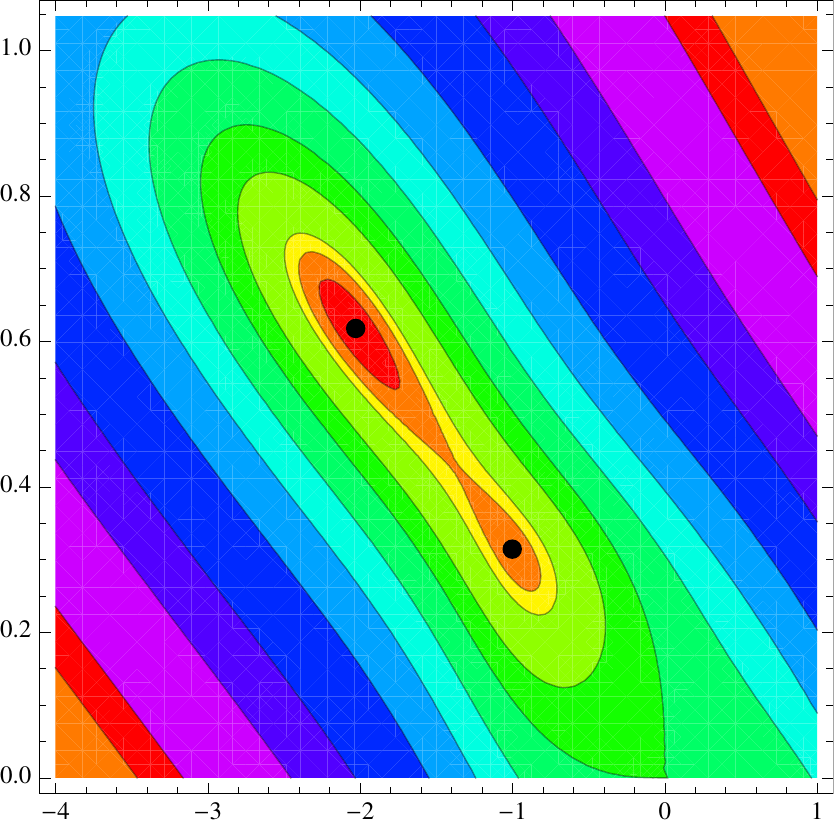} &  
        \includegraphics[width = 0.4\textwidth]{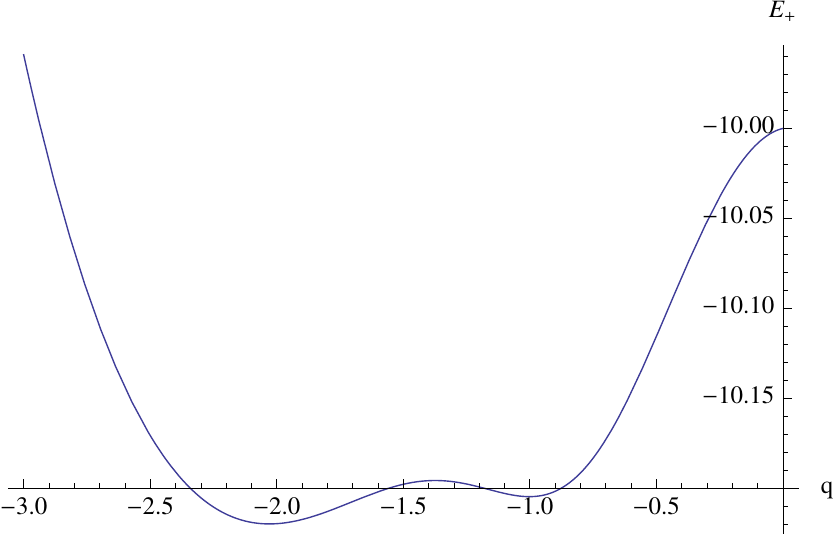}  \\
\begin{tabular}{c}  $\gamma = 0.560$ \\  ~\\ ~ \\ ~ \\  ~\\ ~ \\ ~    \end{tabular} &            \includegraphics[width = 0.4\textwidth]{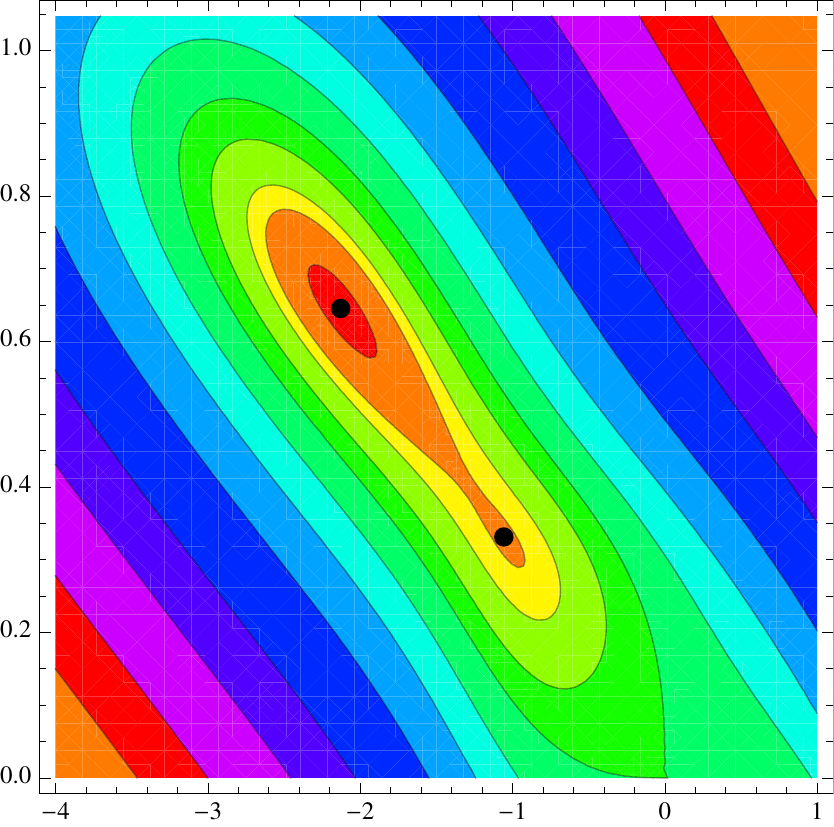} &  
        \includegraphics[width = 0.4\textwidth]{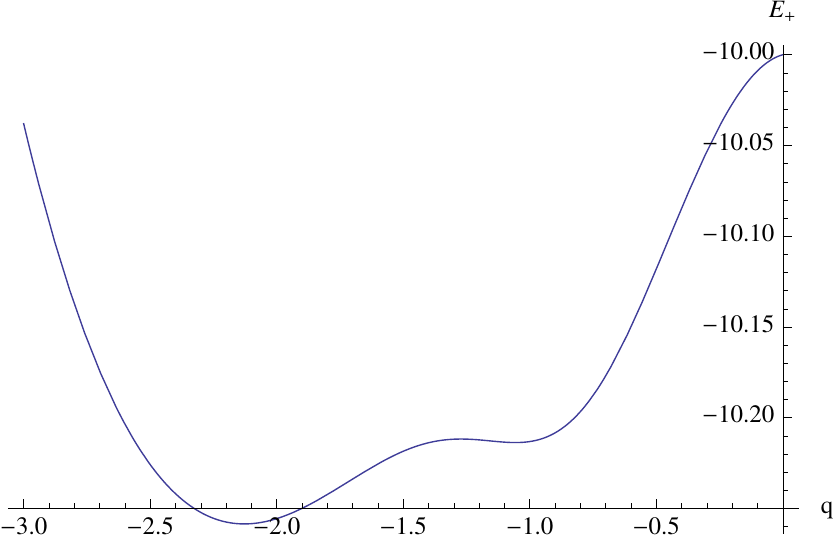}  
\end{tabular}
\vskip -1.5cm
\caption{The symmetry adapted energy surface showing the minima for $\gamma= 0.555, 0.560$ plotted as a function of $q$ and $\theta$ (left) and as function of $q$ (right).}
\label{fig2}
\end{figure}

On the right hand side, the SAES is plotted against $q$, with $\theta$ varying as a linear function of $q$ which crosses both minima. \Fref{fig1} displays the results for $\gamma= 0.545, 0.550, 0.552$, and \fref{fig2}  for $\gamma= 0.555$ and $0.560$. 
It is clearly seen that for $\gamma= 0.545$ the SAES has a global minimum around $q=1$, and a local one around $q=2$ is very shallow. It becomes 
deeper for $\gamma= 0.550$, where the two local minima are well developed, but still the first one is the more profound. At $\gamma= 0.552$ the two minima have the same depth. This is the point where the phase transition takes place. The two cases shown in  \fref{fig2} display the evolution of the absolute minimum around $q=2$.  

The phase transition in the Dicke model, in resonance, for finite number of atoms $N=20$, takes place at a critical value $\gamma_c= 0.552$, 10\% larger than the critical value $0.5$ in the thermodynamic limit. As expected, with increasing $N$ it tends to this limit. The existence of the phase transition for finite values of $N$ has also been found as a singularity in the fidelity susceptibility obtained through exact diagonalization \cite{zanardi,gu}. In the Dicke model they are similar to those found using SACS \cite{Cas12}. There are, however, some important differences. On one hand, the SACS approach provides a description of all observables and of the phase transition for any number of atoms, avoiding the difficulties inherent to the diagonalization. Two order parameters (the number of photons inside the cavity and the fraction of excited atoms) have finite values ($q_c/2$ and $\cos\theta_c$, respectively) in the normal phase, improving in this way the mean field treatment.  
On the other hand, the jump in the critical variables $q_c, \theta_c$ reflects directly in a jump in these order parameters \cite{Cas12}. While it serves as an indicator of the presence of a phase transition at finite $N$, this discontinuity is spurious, in the sense that it is not present in the exact numerical calculations. At the same time, there is a universal relationship between the two order parameters which is exact, and is fully satisfied by the present description \cite{Cas12}.
A detailed analysis of the critical strengths $\gamma_c$ for a finite number of atoms will be presented elsewhere \cite{Cas12b}.

\section{Conclusions}

The phase transition in the Dicke model, for a finite number of atoms $N$, has been presented employing symmetry adapted coherent states. 
Analytic expressions for the symmetry adapted energy surface were presented. It was minimized numerically. It was shown that around the phase transition it has two minima, and the competition between them determines the absolute one. It allows for a precise determination of the critical strength  for any given $N$, which tends to its mean field value in the thermodynamic limit. It provides finite values for the number of photons inside the cavity and the fraction of excited atoms in the normal phase, improving the mean field results, but at the price of having a discontinuity at the phase transition, which is erased in the limit $N \rightarrow \infty$.  The Dicke model keeps offering fascinating new experimental and analytical results.  

This work was partially supported by CONACyT-M\'exico, and DGAPA-UNAM under project IN102811.

\end{document}